\documentclass[%
 reprint,
superscriptaddress,
showpacs,
 amsmath,amssymb,
 aps,
floatfix,
]{revtex4-2}

\usepackage{mwe}
\usepackage{graphicx}
\usepackage{dcolumn}
\usepackage{bm}
\usepackage{color}
\usepackage[normalem]{ulem}
\usepackage{hyperref}
\usepackage{orcidlink}

\begin{document}

\title{Nonlinear Biomechanical Resonances in Birdsong} 

\author{Facundo Fainstein \orcidlink{0000-0002-3700-0106}}
\affiliation{Universidad de Buenos Aires, Facultad de Ciencias Exactas y Naturales, Departamento de Física, Ciudad Universitaria, 1428 Buenos Aires, Argentina.}
\affiliation{CONICET - Universidad de Buenos Aires, Instituto de Física Interdisciplinaria y Aplicada (INFINA), Ciudad Universitaria, 1428 Buenos Aires, Argentina.}

\author{Franz Goller \orcidlink{0000-0001-5333-1987}}
\affiliation{Institute of Integrative Cell Biology and Physiology, University of Münster, Münster 48143, Germany.}
\affiliation{School of Biological Sciences, University of Utah, Salt Lake City, Utah 84112, USA.}

\author{Gabriel B. Mindlin \orcidlink{0000-0002-7808-5708}}
\email[Corresponding author: ]{gabo@df.uba.ar}
\affiliation{Universidad de Buenos Aires, Facultad de Ciencias Exactas y Naturales, Departamento de Física, Ciudad Universitaria, 1428 Buenos Aires, Argentina.}
\affiliation{CONICET - Universidad de Buenos Aires, Instituto de Física Interdisciplinaria y Aplicada (INFINA), Ciudad Universitaria, 1428 Buenos Aires, Argentina.}


\begin{abstract}
Evolution has shaped animal bodies, yet to what extent biomechanical systems impose constraints and provide opportunities across different behaviors remains unclear. In birds, quiet breathing operates at a resonance of the respiratory biomechanics, but song, a behavior thought to be shaped by strong sexual selection, requires much higher breathing rates. Combining physiological recordings with a nonlinear biomechanical model, we show in canaries (\textit{Serinus canaria}) that song production drives the system into a nonlinear regime that broadens the frequency range of amplified responses. This enhancement encompasses the full range of syllabic rates, with an average magnification of $\sim94\%$ of the theoretical maximum. Thus, birds sing at a resonance, indicating that rapid song rhythms evolved to operate under shifting natural frequencies of the respiratory biomechanics. Our results illustrate a shared optimization strategy across behavioral states, reveal a deep connection between neural and biomechanical dynamical parameters and show that sexually selected displays may still rely on optimization strategies.
\end{abstract}

\maketitle


\section{Introduction}
Physical properties of animal bodies are shaped by evolution. A classic example is the modification of energy-harvesting traits such as jaws and beaks \cite{grant2006evolution}. Natural selection has also resulted in efficient execution of fundamental cyclical movements, including locomotion \cite{dickinson1995muscle, dickinson2000animals, tytell2011spikes, kohannim2014analytical, tytell2014role,zhong2021tunable, peleg2025physics} and respiration \cite{crawford1971resonant, kampe1973oscillatory, fainstein2021birds, bates2011oscillation}. However, performance of biomechanical systems is usually assessed for specific scenarios, which fails to account for the diversity of natural behaviors in which they operate. A broader approach is needed to understand how morphological changes constrain behaviors, and to what degree optimization strategies extend across them.

\begin{figure}[t] \centering
    \includegraphics[width=1\linewidth]{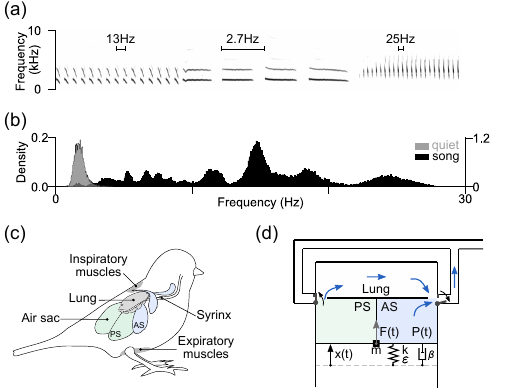} 
    \caption{Avian respiratory system and its timescales across behaviors. (a) Sound spectrogram of a typical canary song composed of trills at different rates, each syllable produced within a respiratory cycle. (b) Distribution of respiratory rates during quiet breathing (in gray, right axis) and song production (in black, left axis). Respiratory frequency during song production expands a broad range, largely departing from quiet breathing rates. Data from five individuals is used (see number of samples in Table S1 in the SM \cite{supplement}). (c) The respiratory system consists of air sacs and a rigid lung. Muscles force the rib cage in or out, thus compressing or expanding the air sacs that act as bellows and ventilate the lung. (d) Elements of the model. A forced piston changes the volume of the cavities representing the posterior (shaded in green, PS) and anterior (shaded in blue, AS) air sacs. Aerodynamic valves direct the flow.}
    \label{fig1}
\end{figure}

We investigated avian respiration in canaries (\textit{Serinus canaria}). Besides its life-sustaining function, the respiratory system plays a central role in vocal production. Songs of many bird species, including canaries, are composed of repeated acoustic elements separated by short silent intervals (vocalizations called trills, Fig. \ref{fig1}(a)). The sound pulses are generated with a high-amplitude expiratory airstream, and the silent intervals are used for short inspirations (mini-breaths) that replenish the air supply \cite{hartley1989airflow}. Control of sound frequency requires coordinated activity of the respiratory and vocal motor systems \cite{amador2008frequency, goguialonso2014motor}, and in canaries, mini-breath trills can reach up to $\sim30Hz$ (Fig. \ref{fig1}(b)). Song production therefore demands delicate and rapid vocal-respiratory motor control and is thought to indicate a singer’s fitness \cite{schmidt2016breathtaking, goller2022vocal}. 

During quiet breathing, respiratory rates match a resonance of the respiratory biomechanics \cite{fainstein2021birds}. This allows an efficient translation of motor instructions into the air-pressure responsible for lung ventilation. However, the respiratory frequency increase during song results in a large departure from the resonance of quiet breathing. Song is a behavior thought to be shaped by strong sexual selection \cite{druaguanoiu2002directional, ballentine2004vocal, nowicki2004song,caro2010female}, that occupies a substantial portion of the animal’s daily activity and entails a measurable direct energetic cost \cite{oberweger2001metabolic, franz2003respiratory, ward2003energy}. Thus, understanding this deviation represents an opportunity to investigate key aspects of behavior, including (i) constrains by physics and physical properties of the body, (ii) the degree of synergistic interaction between the nervous system and the biomechanics, and (iii) the relationship between fitness and optimization strategies in sexually selected displays \cite{zahavi1999handicap}.

\section{Results}
\subsection{Nonlinear biomechanical model}
To address these questions, we first compare respiratory dynamics during quiet breathing and song production. The avian respiratory system consists of a set of air sacs, which are compressed or expanded by expiratory and inspiratory muscles, and these movements ventilate rigid lungs (Fig. \ref{fig1}(c))\cite{maina2025biology}. During inspiration, the thorax expands, drawing air into the posterior air sacs (PS) and through the lungs into the anterior air sacs (AS). During expiration, the air sacs are compressed, such that air from the anterior sacs is exhaled, while air from the posterior sacs passes through the lungs. Aerodynamic valves ensure this unidirectional flow through the lungs, which are perfused with oxygenated air during both respiratory phases. 

An avian respiration model was proposed in terms of three cavities and a piston, incorporating aerodynamic valves, an elastic restitution for the piston, and an external forcing (Fig. \ref{fig1}(d)) \cite{fainstein2021birds}. For small forcing amplitudes, such as those required to reproduce normal breathing, the model was assumed to be linear. Fitting the model to data recorded during quiet breathing revealed that the average respiratory frequency, which is $\sim2Hz$ (Fig. \ref{fig1}(b)), matches the resonance of the model for air pressure, ensuring efficient lung ventilation \cite{fainstein2021birds}.

During singing, the dynamics departs markedly from the one observed during normal respiration. Whereas in quiet breathing the system operates near its linear resonance, the respiratory rhythm during song displays a strikingly different structure: a multimodal distribution with peaks at frequencies that reflect the complex rhythmic composition of the vocalization (see Fig. \ref{fig1}(a-b))\cite{trevisan2006nonlinear, fainstein2025song}. Remarkably, the rate of respiratory cycles can increase by up to $15$ times relative to resting respiration, which is well beyond the resonance identified in the linear regime. In addition, syringeal oscillations are induced by expiratory pressures that are at least an order of magnitude greater than during normal respiration \cite{hartley1989airflow, fainstein2025song, gardner2001simple}.

A nonlinear regime of the biomechanics is engaged during song production. On the one hand, the system is driven in a range of the variables where the restitution force becomes nonlinear \cite{scheid1969volume}. On the other hand, the syringeal lumen narrows during expiration, facilitating high pressurization, and widens during inspiration \cite{suthers1999neuromuscular, goller1996role, goller1996rolephonology}, a behavior that can be mathematically described by a nonlinearity. Is it within this nonlinear regime that a range of frequencies emerges for which an input can evoke an amplified output? 

Applying Newton’s law to the displacement of the piston $x(t)$ and considering the thermodynamics of a cylinder with an opening and gauge pressure $p(t)$, we derived a simple biomechanical model \cite{fainstein2021birds}. Introducing the nonlinearities which are relevant during song production, performing an adiabatic elimination of the inertial term and nondimensionalizing the dynamical variables, the resulting model reads:
\begin{equation}
\tau_{x} \dot{x} = -\left( 1 + x^{2} \right)x - p + F_{0}f(t),    
\end{equation}
\begin{equation}
\tau_{p} \dot{p} =  -\left( 1 + x^{2} \right)x - \left( 1 + \alpha(p) \right) p + F_{0}f(t),
\end{equation}
where $\tau_{x}=\frac{\beta}{k}$, $\tau_{p}=\frac{V_{0}\beta}{P_{0}A^{2}}$, $\alpha_{i,o}=\frac{\tilde{\alpha}_{i,o}\beta}{A^{2}}$, $F_{0}=\frac{f_{0}\varepsilon^{1/2}}{k^{3/2}}$. Here $k$ is the elastic restitution constant, $\beta$ the dissipation coefficient, $A$ the cross-sectional area of the piston, $V_{0}$ the total volume, and $P_{0}$ the atmospheric pressure. The nonlinear contribution to the elastic restitution is modeled with a cubic term of coefficient $\varepsilon$. The relative change in the conductance of the inlet ($\alpha_i$) and outlet ($\alpha_{o}$) flow can be expressed as $\alpha(p)=\alpha_{i}-\left(\alpha_{i}-\alpha_{o} \right)\theta(p)$, with $\theta(p)$ the Heaviside function. The amplitude-normalized function $f(t)$ represents the external forcing, whose amplitude is set by the coefficient $F_{0}$. 
\subsection{Measurement of the variables and parameter fitting}
We collected physiological data from five naturally behaving individuals to measure the dynamic variables and determine the parameters of the model (see Materials and Methods). In these experiments, we recorded air sac pressure and electromiographic activity (EMG) of expiratory muscles. In three other birds, thoracic displacement, a proxy of the displacement of the air sacs, was measured \cite{fainstein2021birds}. During singing, our experiments indicated a strong nonlinearity in the restitution force: the amplitude of thoracic displacement increased by only a factor of two, while air pressure by a factor of fifteen. Finaly, we fitted the model parameters to accurately reproduce the morphology of the pressure gestures when driven by the measured EMG signal, while qualitatively capturing the scaling relationship between the displacement and the pressure (see Materials and Methods). 

\begin{figure*}[t]
  \centering
  \includegraphics[width=\textwidth]{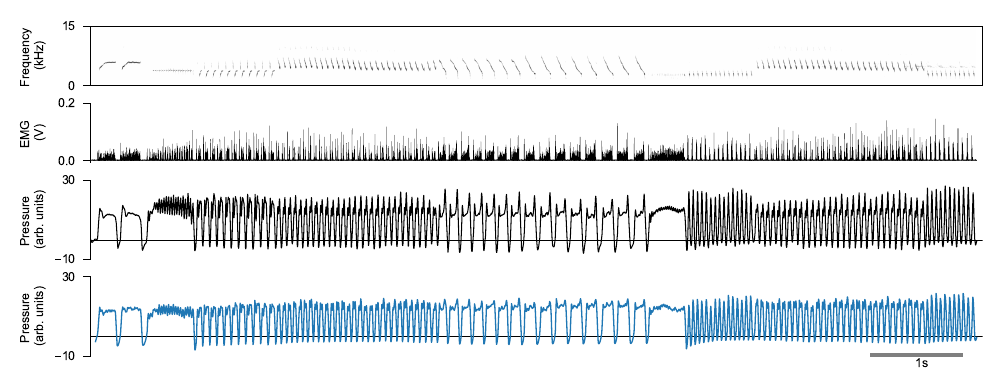}
  \caption{Respiratory dynamics during song production. Model prediction (blue line) of measured air sac pressure patterns (black line) using recorded expiratory electromyographic activity (EMG) as input. Pressure units are in terms of quiet respiration amplitude, which is normalized to $[-1, 1]$. Top panel shows the spectrogram of the recorded sound.}
  \label{fig2}
\end{figure*}

In the first three panels of Fig. \ref{fig2}, we present the simultaneous recordings of sound, EMG activity, and air sac pressure during singing. The fourth panel shows the output of the fitted model. The similarity between the measured and predicted pressure signals is remarkable (Fig. S1 in the Supplemental Material (SM) shows the result for the other individuals \cite{supplement}). The model does not only capture the syllabic rhythm but also accurately reproduces the morphology of individual syllables. A single set of parameters can account for the wide diversity of syllables sung.

\subsection{Resonances and magnification curves}
The frequency response of a linear system is obtained by computing the output amplitude for different harmonic components. The curve that describes the fraction of the output amplitude relative to the maximally possible value is called the magnification curve \cite{feynman1966lectures}. In a nonlinear system, the situation is more subtle, because the response depends on the amplitude of the driving force. In our case, the value of the amplitude is informed by the fit to experimental data. It determines the contribution of the nonlinearities due to tissue restitution and to the difference between the output opening and the input opening to the air sacs—in other words, the two elements that introduce nonlinearities into the model. Once the amplitude is fixed, there exists a range of frequencies within which entrainment occurs and for which magnification can be high.

\begin{figure}[t] \centering
    \includegraphics[width=1\linewidth]{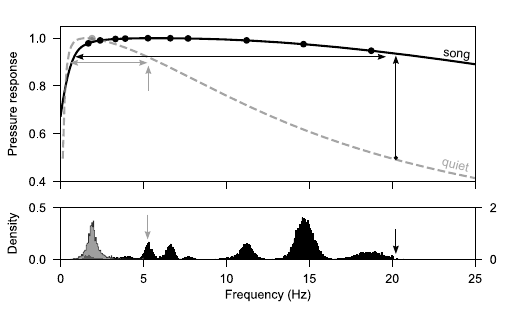} 
    \caption{Birds breathe and sing at resonances of the biomechanics. Magnification curves for quiet breathing (dashed gray) and singing (solid black) reveal a nonlinear enhancement of the bandwidth during song production. Black dots indicate the modal production rate of each syllable type (see Materials and Methods and Fig. S3 in the SM \cite{supplement}). The bottom panel displays the marginal distribution of syllable rates during singing (black, left axis) and the distribution of quiet breathing rates (gray, right axis). Number of samples are displayed in Table S1 in the SM \cite{supplement}. Black arrows in the lower panel indicate the highest syllabic rate observed, which defines a magnification level (marked by a black arrow in the top panel). This magnification level sets the bandwidth, shown by horizontal arrows.}
    \label{fig3}
\end{figure}

To measure the magnification curves, we proceeded as follows. First, for the birds in the study, we fitted the parameters (including the amplitude of the driving force) that allowed us to reproduce the air sac pressures recorded experimentally (see Materials and Methods, Fig. 2, and Figs. S1-S2 in the SM \cite{supplement}). Then, we verified that, for smaller forcing amplitudes, it was possible to reproduce quiet breathing patterns (see Materials and Methods). With each set of parameter pairs, we explored the system’s responses to harmonic forcing and plotted the magnifications for both cases. The curves are shown, for one of the birds in our study, in Fig. \ref{fig3} (the curves corresponding to all other birds are shown in Fig. S4 in the SM \cite{supplement}).

The dashed line shows the system's response in the quiet breathing (linear) regime, while the solid curve corresponds to the response in the singing (nonlinear) regime. The main observation is that the range of amplified frequencies is substantially broader in the nonlinear regime. We find that the bandwidth increases by a factor of approximately $5$ $\left(5.4 \pm 1.5\right)$, averaged across five birds, as defined by the amplification threshold encompassing all sung syllables (indicated with arrows in Fig. 3). This nonlinear enhancement enables the syllables to be produced, on average, at $\sim94\%$ $(94.22 \pm 0.03)$ of the theoretically maximal magnification. In contrast, under linear response conditions, the system would reach only $\sim55\%$ $(55.3 \pm 0.1)$, averaged across individuals. These findings indicate that the system operates at a biomechanical resonance during normal breathing, and that the nonlinear response engaged during singing allows all syllables to be produced with a remarkably high level of magnification.
\section{Discussion}
Behavior emerges from the interaction between the nervous system, the biomechanics and the environment \cite{chiel1997brain, gomez2019life}. A striking way in which this interaction synergistically manifests through the concept of resonance \cite{tytell2011spikes}. In these cases, an input (such as one generated by the nervous system) that is tuned to a particular frequency (in the case of a linear system) or to a range of frequencies (in nonlinear systems) relevant to the biomechanics will have its impact amplified. We have shown that in avian respiration, the bandwidth expansion that arises from the nonlinearities is sufficient to cover all the syllable production rates used during singing, with an average magnification of $\sim94\%$ of the maximally possible. This implies that birds rely on a resonant range of their respiratory biomechanics not only during quiet breathing, but also during song production. These findings reveal how a single biomechanical system can provide optimized performances across behavioral states, with nonlinear resonance as the underlying mechanism, and quantitatively demonstrate an intimate relationship between the dynamics of neural motor circuits and the physical properties of the body and its environment.

The facilitation of a broad range of respiratory rates through nonlinearities in the pressure generating mechanism likely represents a more general principle for how selection has shaped biomechanical systems to ensure optimized execution of diverse behaviors. Importantly, this optimization does not only apply to life-sustaining movements, but extends to sexually selected display movements, such as song production or integration of multimodal displays \cite{franz2004multimodal}. Sexually selected signals are thought to provide honest information about the displaying individual and therefore performance should be subject to handicaps \cite{zahavi1999handicap, gil2002honesty}. The strong indication that respiratory movements for extremely rapid vocal performance still rely on favorable biomechanical features illustrates that assessment of performance in behavior requires careful analysis of all interacting systems \cite{goller2022vocal, ballentine2004vocal, nowicki2004song, podos2009vocal, podos2020vocal}.

\section{Materials and Methods}\label{materiaslandmethods}
\subsection{Experiments}
\subsubsection{Subjects}\label{secsubjects}
Experiments were conducted on five male canaries (\textit{Serinus canaria}), according to the regulation of the animal care committee of the University of Buenos Aires. Birds were acquired from local breeders. After 10-d quarantine, birds were housed individually in $27$ cm $\times$ $23$ cm $\times$ $20$ cm wire cages inside an acoustic chamber. The chamber's door was usually opened, and the subjects had intermittent (but daily) auditory and visual contact with other canaries. They were fed with food and water ad libitum. The photoperiod was set from 6:00AM to 8:00 PM and was controlled by an automatic timer. Measurements were done between October and March (South American spring and summer).

\subsubsection{Surgery}\label{secsurgery}
Surgery was performed under general anesthesia using isoflurane. During surgery, a flexible silicone cannula (AMsystems Silicon Tubing 0.030’’ $\times$ 0.065’’ $\times$ 0.0177’’ Catalog No. 807000) was inserted right below the last rib, 5 mm into the anterior thoracic air sac. The cannula was sutured to the rib cage, and the incision site was sealed with tissue cement (Vetbond; 3 M Animal Care Products, St. Paul, MN, USA). The free end of the cannula was attached to a piezoresistive pressure transducer (FHM-02 PGR, Fujikura, Tokyo, Japan), positioned in the backpack. In five individuals, differential EMG activity was recorded with a pair of fine ($< 0.01$ cm diameter), custom-built, silver-plated copper wires (44TDQ, Phoenix Wire Inc., VT, USA). To record from expiratory abdominal muscles, the bare wire tips (0.2 cm length) of the electrodes were aligned in parallel and inserted into the thin muscle sheet and fixed in place by tissue adhesive (Vetbond) applied to the muscle surface. Insulated leads were routed under the skin to the birds’ back and connected to the backpack from which stronger wires led to the recording unit. Methods used to measure thoracic displacement are described in ref. \cite{fainstein2021birds}.

\subsubsection{Experimental recordings}\label{secexperimental}
Song was recorded using a directional microphone (Venetian HT-81 A) and audio amplifier (Behringer MIC100). All signals were acquired using a National Instruments acquisition board (NIDAQ-USB-6212 or NIDAQ-USB-6259). Recordings were controlled using a custom MATLAB script at $44.15$ kHz sample rate. Pressure signal was connected, before digitization, to a custom-built variable gain analog amplifier ($300 \times$ maximum gain). EMG signals were processed with a 150 Hz high-pass RC filter and with an analog differential amplifier ($225 \times$), both printed on the backpack. All custom-built analog devices were powered by external 12-V batteries to avoid line noise.

\subsection{Model fitting}\label{secmodelfit}
\subsubsection{Muscle forcing}\label{secforcing}
The external drive was determined from the measured expiratory EMG. A critically damped linear oscillator was used to process the EMG signal and approximate the expiratory force generated by the muscles \cite{shapiro2000control, doppler2018electromyographic}, namely:
\begin{equation}
\dot{x}_{1} = x_{2},    
\end{equation}
\begin{equation}
\dot{x}_{2} =-\frac{1}{\tau^{2}}\left( x_{1} - m(t) \right)-\frac{2}{\tau}x_{2}.
\end{equation}
Where $m(t)$ is the absolute value of the EMG. The temporal scale $\tau$ is set to $5ms$, consistent with \textit{in vitro} muscle stimulation of these muscles \cite{srivastava2017motor}. The normalized expiratory force is obtained as $f_{exp}(t)=\frac{x_{1}(t)-P_{10}}{P_{90}-P_{10}}$, with $P_{10}$ and $P_{90}$ the $10^{th}$ and $90^{th}$ percentiles of $x_{1}$, respectively. A threshold of three times the standard deviation of $x_{1}$, computed in an interval in which there was no sound produced, is used to determine the intervals of non-activity of the expiratory muscles. A half-wave sinusoidal function approximates the inspiratory force, with non-zero values in each of these intervales ($f_{insp}(t)$). The amplitude-normalized external force is obtained by the summation $f(t)=f_{exp}(t)+f_{insp}(t)$.

\subsubsection{Fitting procedure}\label{secmodelfit}
Parameter fitting was performed by minimizing the squared distance between the pressure obtained by integrating the model and the measured pressure. Pressure segments recorded during singing were normalized such that quiet respiration oscillates in the range $\left [ -1 , 1 \right ]$. With this normalization, the model is fitted to reproduce both the measured pressure signal and the relative amplitude changes between the two regimes. A grid search fitting procedure with $\geq 4.9 \times 10^{4}$ grid points was performed over the parameter intervals shown in Table S2 in the SM \cite{supplement}. Once the best-fit parameters were obtained, quiet respiration data was fitted as follows. Five consecutive cycles of quiet breathing were selected. Following ref. \cite{fainstein2021birds}, a square-wave force with normalized amplitude was used. Parameters not expected to vary between regimes ($\tau_{x}$ and $\tau_{p}$) were held fixed, and equal flow conductance was assumed for inspiration and expiration ($\alpha_{i}=\alpha_{o}$). A grid search with $10^{4}$ points over the $\left(\alpha_{i}, F_{0} \right)$ parameter space was performed to estimate the minimum squared distance between the modeled and measured pressure. The parameters obtained for each regime and for each of the five birds analyzed are shown in the Fig. S2 in the SM \cite{supplement}.

\subsection{Song syllabic timescales}\label{secclustering}
To analyze the distribution of syllabic rates (defined as the inverse of syllable duration) within each syllable type, we performed an automatic clustering of air sac pressure patterns as follows. First, syllables were segmented from the recordings, resampled to $1000$ points, and normalized to the range $\left [ -1 , 1 \right ]$. For each bird, all resampled and normalized syllables were arranged into an $N_{s} \times 1000$ matrix, where $N_{s}$ is the number of syllables for that bird. We then applied a two-dimensional Uniform Manifold Approximation and Projection (UMAP, using umap-learn for Python), which embedded each syllable pressure pattern into a two-dimensional space, revealing clearly separated clusters. These clusters were identified algorithmically using Hierarchical Density-Based Spatial Clustering of Applications with Noise (HDBSCAN), with a minimum cluster size set to $2\%$ of the dataset (via scikit-learn)\cite{campello2013density}. Fig. S3 in the SM shows the result of this procedure for bird 1 \cite{supplement}.

\begin{acknowledgments}
We are grateful to A. Amador, S. M. Geli, J. F. Döppler, R. Bistel, F. L. Leites, and P. Clark Di Leoni for valuable help and discussions. This work was partially funded by ANPCyT-FONCyT (Argentina) under Grant PICT-2018-00619, Universidad de Buenos Aires (UBACyT, Argentina), and Consejo Nacional de Investigaciones Científicas y Técnicas (CONICET, Argentina).
\end{acknowledgments}

\section*{DATA AVAILABILITY}
All the data and codes that support the findings of this study are publicly available in Zenodo \cite{zenodo}.

\bibliography{references}

\end{document}


\title{Supplemental Material for Nonlinear Biomechanical Resonances in Birdsong}
\author{Facundo Fainstein \orcidlink{0000-0002-3700-0106}}
\affiliation{Universidad de Buenos Aires, Facultad de Ciencias Exactas y Naturales, Departamento de Física, Ciudad Universitaria, 1428 Buenos Aires, Argentina.}
\affiliation{CONICET - Universidad de Buenos Aires, Instituto de Física Interdisciplinaria y Aplicada (INFINA), Ciudad Universitaria, 1428 Buenos Aires, Argentina.}

\author{Franz Goller \orcidlink{0000-0001-5333-1987}}
\affiliation{Institute of Integrative Cell Biology and Physiology, University of Münster, Münster 48143, Germany.}
\affiliation{School of Biological Sciences, University of Utah, Salt Lake City, Utah 84112, USA.}

\author{Gabriel B. Mindlin \orcidlink{0000-0002-7808-5708}}
\email[Corresponding author: ]{gabo@df.uba.ar}
\affiliation{Universidad de Buenos Aires, Facultad de Ciencias Exactas y Naturales, Departamento de Física, Ciudad Universitaria, 1428 Buenos Aires, Argentina.}
\affiliation{CONICET - Universidad de Buenos Aires, Instituto de Física Interdisciplinaria y Aplicada (INFINA), Ciudad Universitaria, 1428 Buenos Aires, Argentina.}


\maketitle
\onecolumngrid

\begin{figure}[!t]
\centering
\includegraphics[width=\textwidth]{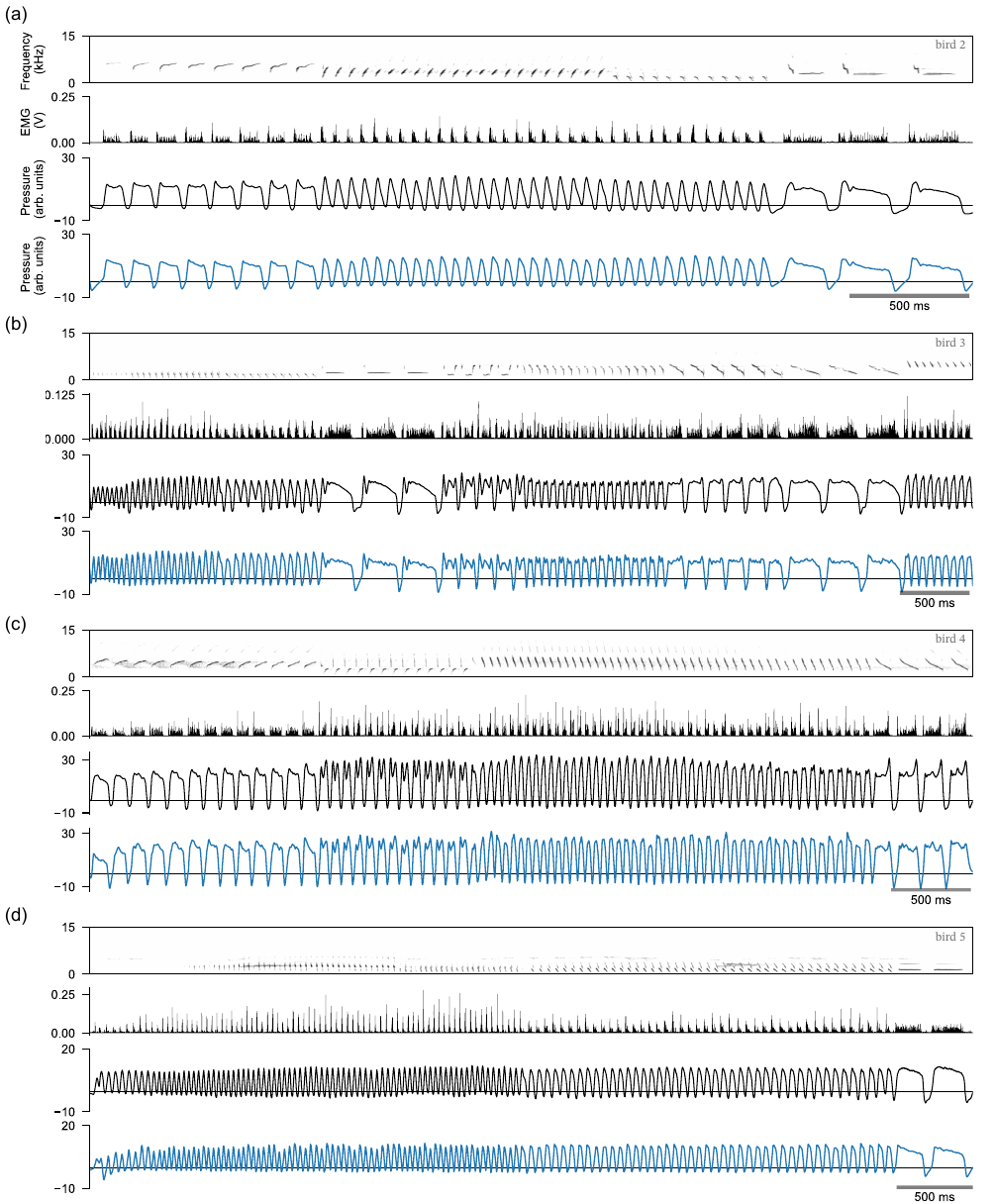}
\caption{Biomechanical dynamics during song production. Model prediction (blue line) of measured air sac pressure patterns (black line) using expiratory electromyographic activity (EMG) as input, for bird 2 (a), bird 3 (b), bird 4 (c), and bird 5 (d). Spectrogram of the recorded sound is included in each top panel.}
\label{figS_fits}
\end{figure}

\begin{figure}[!t]
\centering
\includegraphics[width=\textwidth]{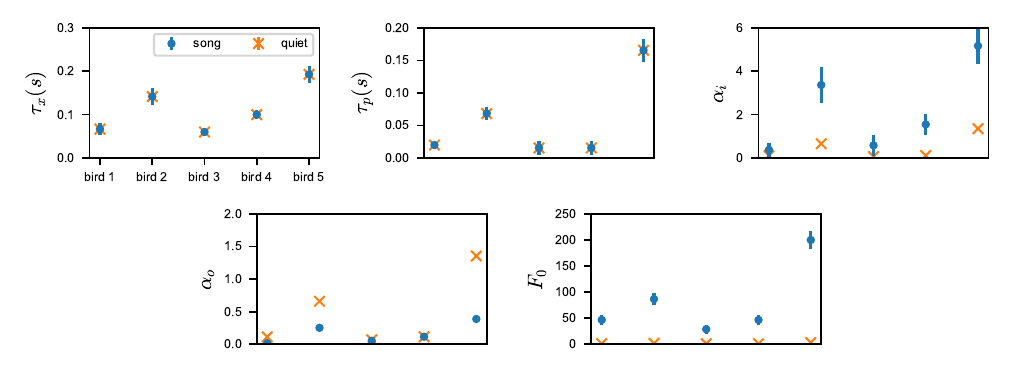}
\caption{Best-fit parameter values for each of five individuals, fitting song production (dots) and quiet breathing (crosses). Bars indicate the parameter-grid spacing used in the search.}
\label{figS_pars}
\end{figure}

\begin{figure}[!t]
\centering
\includegraphics[width=.5\textwidth]{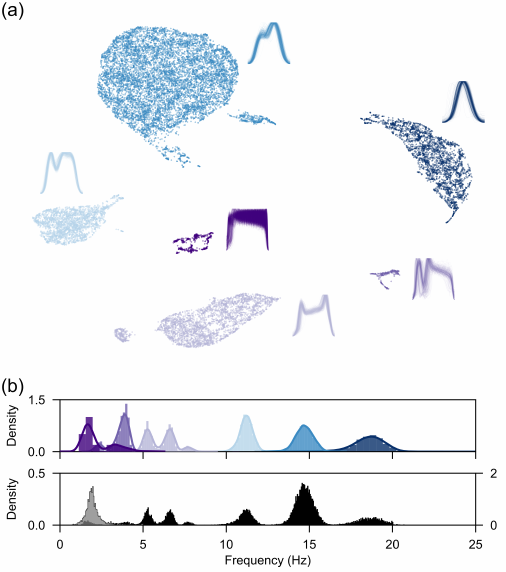}
\caption{Pressure patterns clustering and modal syllabic rates. (a) Two-dimensional UMAP projection to cluster air sac pressure patterns (bird 1). (b) Conditional distributions of syllabic rates given syllable type (top panel), following the color-code of the pressure pattern clusters. These distributions are used to determine the modal production rate of each syllable type (black dots in Fig. 3). The bottom panel displays the marginal distribution of syllable rates during singing (black, left axis) and the distribution of quiet breathing rates (gray, right axis).}
\label{figS_clustering}
\end{figure}

\begin{figure}[!t]
\centering
\includegraphics[width=\textwidth]{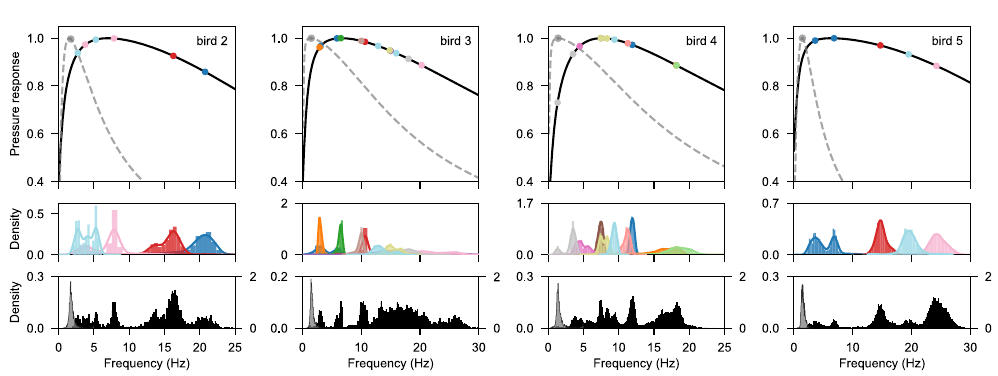}
\caption{Birds breathe and sing at a resonance. Magnification curves for quiet breathing (gray dashed lines) and song production (black lines), for the birds not shown in the main manuscript. For each bird, the modal syllable rates of each song are indicated by color-coded dots, corresponding to the conditional distributions shown in the middle panel. The bottom panel displays the marginal distribution of syllable rates during singing (black, left axis) and the distribution of quiet breathing rates (gray, right axis).}
\label{figS_resonances}
\end{figure}

\clearpage
\begin{table}[t]
\caption{Number of samples per individual in each behavioral regime. These samples are used to construct the histograms shown in Fig. 1, Fig. 3, and Figs. S3-4.}
\centering
\setlength{\tabcolsep}{12pt}
\begin{tabular}{lcc}
\hline\hline
Bird ID & Number of song syllables & Number of quiet respiratory cycles \\
\hline
CaFF016-VioVio (bird 1) & 19105 & 6610 \\
CaFF-NeVe (bird 2) & 2907 & 3189 \\
CaFF-BlaVe (bird 3) & 3456 & 2818 \\
CaFF073-RoVio (bird 4) & 16464 & 2646 \\
CaFF909-NaRo (bird 5) & 15336 & 2361 \\
\hline\hline
\end{tabular}
\label{tab:bird_data}
\end{table}

\begin{table}[t]
\caption{Parameter intervals used to perform a grid-search fitting procedure of song segments.}
\centering
\setlength{\tabcolsep}{12pt} 
\begin{tabular}{lc}
\hline\hline
Parameter & Interval \\
\hline
$\tau_x$ (s) & $[0.01, 0.25]$ \\
$\tau_p$ (s) & $[0.005, 0.2]$ \\
$\alpha_i$ & $[0.05, 6]$ \\
$\alpha_{\mathrm{rel}} = \alpha_i / \alpha_o$ & $[0.025, 0.125]$ \\
$F_0$ & $[20, 200]$ \\
\hline\hline
\end{tabular}
\label{tab:parameter_intervals}
\end{table}